\documentclass[]{article}
\begin{document}
\title{Invariants and polynomial identities for higher rank matrices}

\author{{\bf Victor Tapia}\\
\\
{\it Departamento de Matem\'aticas}\\
{\it Universidad Nacional de Colombia}\\
{\it Bogot\'a, Colombia}\\
\\
{\tt tapiens@gmail.com}}

\maketitle

\begin{abstract}

We exhibit explicit expressions, in terms of components, of discriminants, determinants, characteristic polynomials and polynomial identities for matrices of higher rank. We define permutation tensors and in term of them we construct discriminants and the determinant as the discriminant of order $d$, where $d$ is the dimension of the matrix. The characteristic polynomials and the Cayley--Hamilton theorem for higher rank matrices are obtained there from.

\end{abstract}

\maketitle

\section{Introduction}

A matrix ${\bf A}$ of rank $r$ is an array of numbers $A_{i_1\cdots i_r}$ where the indices $i$ run from $1$ to $d$, the dimension of the matrix. 

For $r=2$ we have the ordinary matrices ${\bf a}$, that is, square arrays of numbers $a_{ij}$. In this case it is possible to define a matrix addition and a matrix multiplication. Then, according to the Cayley--Hamilton theorem, only the first $d$ powers of ${\bf a}$ are linearly independent. The traces of these $d$ powers are a set of invariants which help to characterise the matrix ${\bf a}$ and several of its properties.

For $r>2$ the situation is different since, even when there is a matrix addition, a natural matrix multiplication does not exist. The absence of a natural multiplication operation makes difficult to introduce concepts analogous to those which are standard in matrix calculus, namely, traces, invariants, polynomials, etc. Therefore, for matrices of higher rank there is not a result similar to the Cayley--Hamilton theorem. Furthermore, in principle, we have neither a way of determining the number of algebraically independent invariants (if any), nor a way to construct polynomial identities.
 
Several results related to invariants and polynomial identities for matrices of higher rank are known \cite{GKZ2,GKZ3}. However, in general, explicit expressions in terms of the components $A_{i_1\cdots i_r}$ are lacking. 

The purpose of this article is to show that, in spite of the absence of a matrix multiplication, 
it is possible to define invariants, in fact a finite number $d$, that it is possible to define a determinant, and that there exists a polynomial identity similar to the statement of the Cayley--Hamilton theorem. 

In section 2 we start with a reminder of some standard results for ordinary matrices. In this case, invariants can be obtained, for example, as traces of the powers of the matrix ${\bf a}$. Other family of invariants are the discriminants. Discriminants are a more convenient set of invariants than traces since only the first $d$ of them are non trivial, while the rest are identically zero. Particularly interesting is the discriminant of order $d$ which corresponds to the determinant of ${\bf a}$. The Cayley--Hamilton theorem is formulated as the condition for the vanishing of a certain polynomial relation among discriminants and powers of the matrix ${\bf a}$. Next, we introduce an index notation similar to that of tensor calculus and all matrices are then treated as tensors. Discriminants can be defined in terms of permutation products with a metric tensor ${\bf g}$. Since, this second definition makes no reference to traces, this definition of discriminants is best suited than traces for the generalization to higher rank matrices.

In section 3 we consider higher rank matrices. Since there is not a concept similar to that of a multiplication operation, the construction of invariants must be done by generalizing  definitions which are independent of the concept of multiplication. For instance, discriminants can be generalized to higher rank matrices if we adopt the definition based on permutation products. The even rank and the odd rank cases must be studied separately. For even rank matrices the fourth--rank matrices are the prototype. By using permutation products we construct the corresponding discriminants and define the determinant as the discriminant of order $d$. Then we show that a certain polynomial relation among discriminants and products of the matrix ${\bf A}$ vanishes identically, which is a statement similar to the Cayley--Hamilton theorem. For odd rank matrices, a naive generalization of the results above leads to useless relations. Instead, we show that in this case it is necessary to introduce an even rank matrix as the direct product of the original odd rank matrix and construct discriminants and polynomial identities as for the even rank case.

Section 4 is dedicated to the conclusions. For the sake of simplicity in the present work we restrict our considerations to completely symmetric matrices and tensors. Generalizations to other situations (other ranks and more general symmetries) are easy to implement, but we refrain to exhibit them for they do not add any new understanding to the problem.

Some preliminary and imcomplete results, similar to the ones reported here, can be found in \cite{Ta05}. In that work our emphasis was on the construction of invariants using a graphical algorithm based on semi--magic squares. However, the rigorous mathematical justification for that algorithm are the invariants constructed with matricial properties alone as exhibited here.

\section{Preliminaries}

Due to the nature of our approach, we must regretfully bore the reader by exhibiting some standard and well known results that are needed to show how the concepts of invariant, discriminant, determinant, characteristic polynomials, polynomial identities, and other concepts of matrix calculus, correctly generalize to matrices of a higher rank. Even when these results can be found in standard references, see for example \cite{GKZ3}, we include them here because we need them written in very particular forms (which are not to be found in a unique reference) which make easier, even conceptually, the passage to matrices of higher rank.

\subsection{Basic definitions and results}

A matrix ${\bf a}$ is a square array of numbers $a_{ij}$, with $i,j=1,\cdots,d$, where $d$ is the dimension of the matrix. Let ${\bf b}$ be a second matrix with components $b_{ij}$. Then, the matrix multiplication is defined by the product matrix ${\bf c}={\bf a}\cdot{\bf b}$ with components

\begin{equation}
c_{ij}=\sum_{k=1}^d\,a_{ik}\,b_{kj}\,.
\label{01}
\end{equation}

\noindent This operation is the Cartesian product among rows and columns. The unit element ${\bf I}$ for the matrix multiplication, ${\bf a}\cdot{\bf I}={\bf I}\cdot{\bf a}={\bf a}$, is the matrix with components $I_{ij}$ given by

\begin{equation}
I_{ij}=\cases{1\,,&for $i=j$;\cr0\,,&otherwise.\cr}
\label{02}
\end{equation}

The determinant of ${\bf a}$, $\det({\bf a})$, given by the Leibniz formula

\begin{equation}
\det({\bf a})=\sum_{\pi\in S_d}\,{\rm sign}(\pi)\,\left(\prod_{i=1}^d\,a_{i\pi(i)}\right)\,.
\label{03}
\end{equation}

\noindent If the determinant (\ref{03}) is different from zero, then there exist an inverse matrix ${\bf a}^{-1}$ satisfying ${\bf a}^{-1}\cdot{\bf a}={\bf a}\cdot{\bf a}^{-1}={\bf I}$. In terms of components we have

\begin{equation}
\sum_{k=1}^d\,(a^{-1})_{ik}\,a_{kj}=\sum_{k=1}^d\,a_{ik}\,(a^{-1})_{kj}=I_{ij}\,.
\label{04}
\end{equation}

\noindent Then the inverse matrix is given by

\begin{equation}
{\bf a}^{-1}={{{\rm adj}({\bf a})}\over{\det({\bf a})}}\,,
\label{05}
\end{equation}

\noindent where ${\rm adj}({\bf a})$ is the adjoint matrix.

The product of a matrix ${\bf a}$ with itself, that is ${\bf a}^2$, is the matrix with components

\begin{equation}
({\bf a}^2)_{ij}=\sum_{k=1}^d\,a_{ik}\,a_{kj}\,.
\label{06}
\end{equation} 

\noindent Powers of ${\bf a}$ of an order $s$, ${\bf a}^s$, are the matrices with components

\begin{equation}
({\bf a}^s)_{ij}=\underbrace{\sum_{k_1=1}^d\,\cdots\,\sum_{k_{s-1}=1}^d}_{s-1\,{\rm times}}\,
\underbrace{a_{ik_1}\,\cdots\,a_{k_{s-1}j}}_{s\,{\rm times}}\,.
\label{07}
\end{equation}

\noindent By definition ${\bf a}^0={\bf I}$ and ${\bf a}^1={\bf a}$. The trace of a matrix ${\bf a}$ is given by

\begin{equation}
{\bf tr}({\bf a})=\sum_{i=1}^d\,a_{ii}\,.
\label{08}
\end{equation}

\noindent The trace of ${\bf a}^s$ is given by

\begin{equation}
{\bf tr}({\bf a}^s)=\sum_{i=1}^d\,({\bf a}^s)_{ii}\,.
\label{09}
\end{equation}

\noindent Furthermore, ${\bf tr}({\bf a}^0)=d$.

Let us now consider the eigenvalue problem

\begin{equation}
[{\bf a}-\lambda\,{\bf I}]\,{\bf v}=0\,.
\label{10}
\end{equation}

\noindent The condition for the existence of a solution is

\begin{equation}
\det[{\bf a}-\lambda\,{\bf I}]=0\,.
\label{11}
\end{equation}

\noindent From here it follows that the eigenvalues are related to the traces by

\begin{equation}
{\bf tr}({\bf a}^s)=\sum_{k=1}^d\,\lambda^s_k\,.
\label{12}
\end{equation}

An invariant is a quantity which does not change under similarity transformations. Invariants are important because they help to characterise a matrix ${\bf a}$ and several of its properties. Examples of invariants are the traces (\ref{09}). 

Discriminants are another family of invariants which are constructed as follows. Let us consider a vector ${\vec\lambda}=\lambda_1,\lambda_2,\cdots$, with an infinite number of components. Then, the elementary symmetric products ${\bf P}$ are defined by

\begin{equation}
P_s({\vec\lambda})=\sum_{k_1\not=\cdots\not=k_s}\,\lambda_{k_1}\,\cdots\,\lambda_{k_s}\,.
\label{13}
\end{equation}

\noindent The power sums ${\bf Q}$ are defined by

\begin{equation}
Q_s({\vec\lambda})=\sum_{k=1}^\infty\,\lambda_k^s\,.
\label{14}
\end{equation}

\noindent Elementary symmetric products and power sums are related by

\begin{eqnarray}
P_0&=&1\,,\nonumber\\
P_1&=&Q_1\,,\nonumber\\
P_2&=&{1\over2}\,\left[Q_1^2-Q_2\right]\,,\nonumber\\
P_3&=&{1\over{3!}}\,\left[Q_1^3-3\,Q_1\,Q_2+2\,Q_3\right]\,,\nonumber\\
P_4&=&{1\over{4!}}\,\left[Q_1^4-6\,Q_1^2\,Q_2+8\,Q_1\,Q_3+3\,Q_2^2-6\,Q_4\right]\,,\nonumber\\
P_5&=&{1\over{5!}}\,\left[Q_1^5-10\,Q_1^3\,Q_2+15\,Q_1\,Q_2^2+20\,Q_1^2\,Q_3\right.\nonumber\\
&&\qquad\left.-20\,Q_2\,Q_3-30\,Q_1\,Q_4+24\,Q_5\right]\,,\nonumber\\
P_6&=&{1\over{6!}}\,\left[Q_1^6-15\,Q_1^4\,Q_2+45\,Q_1^2\,Q_2^2-15\,Q_2^3\right.\nonumber\\
&&\qquad+40\,Q_1^3\,Q_3-120\,Q_1\,Q_2\,Q_3+40\,Q_3^2\nonumber\\
&&\qquad\left.-90\,Q_1^2\,Q_4+90\,Q_2\,Q_4+144\,Q_1\,Q_5-120\,Q_6\right]\,,
\label{15}
\end{eqnarray}

\noindent These are the Newton relations.

If only the first $d$ components of ${\vec\lambda}$ are different from zero, then $P_s\equiv0$ for $s>d$.

\noindent Therefore, traces are in correspondence with elementary symmetric products. In order to complete the correspondence, and following relations (\ref{15}), the discriminants of a matrix ${\bf a}$ are defined by

\begin{eqnarray}
c_0({\bf a})&=&1\,,\nonumber\\
c_1({\bf a})&=&\bigl<{\bf a}\bigr>\,,\nonumber\\
c_2({\bf a})&=&{1\over2}\,\left[\bigl<{\bf a}\bigr>^2-\bigl<{\bf a}^2\bigr>\right]\,,\nonumber\\
c_3({\bf a})&=&{1\over{3!}}\,\left[\bigl<{\bf a}\bigr>^3-3\,\bigl<{\bf a}\bigr>\,\bigl<{\bf a}^2
\bigr>+2\,\bigl<{\bf a}^3\bigr>\right]\,,\nonumber\\
c_4({\bf a})&=&{1\over{4!}}\,\left[\bigl<{\bf a}\bigr>^4-6\,\bigl<{\bf a}\bigr>^2\,\bigl<{\bf a}
^2\bigr>+8\,\bigl<{\bf a}\bigr>\,\bigl<{\bf a}^3\bigr>+3\,\bigl<{\bf a}^2\bigr>^2-6\,\bigl<{\bf a}^4\bigr>\right]\,,\nonumber\\
c_5({\bf a})&=&{1\over{5!}}\,\left[\bigl<{\bf a}\bigr>^5-10\,\bigl<{\bf a}\bigr>^3\,\bigl<{\bf a}
^2\bigr>+15\,\bigl<{\bf a}\bigr>\,\bigl<{\bf a}^2\bigr>^2+20\,\bigl<{\bf a}\bigr>^2\,\bigl<{\bf a}^3\bigr>\right.\nonumber\\
&&\qquad\left.-20\,\bigl<{\bf a}^2\bigr>\,\bigl<{\bf a}^3\bigr>-30\,\bigl<{\bf a}\bigr>\,\bigl<
{\bf a}^4\bigr>+24\,\bigl<{\bf a}^5\bigr>\right]\,,\nonumber\\
c_6({\bf a})&=&{1\over{6!}}\,\left[\bigl<{\bf a}\bigr>^6-15\,\bigl<{\bf a}\bigr>^4\,\bigl<{\bf a}
^2\bigr>+45\,\bigl<{\bf a}\bigr>^2\,\bigl<{\bf a}^2\bigr>^2-15\,\bigl<{\bf a}^2\bigr>^3\right.
\nonumber\\
&&\qquad+40\,\bigl<{\bf a}\bigr>^3\,\bigl<{\bf a}^3\bigr>-120\,\bigl<{\bf a}\bigr>\,\bigl<{\bf a}
^2\bigr>\,\bigl<{\bf a}^3\bigr>+40\,\bigl<{\bf a}^3\bigr>^2\nonumber\\
&&\qquad-90\,\bigl<{\bf a}\bigr>^2\,\bigl<{\bf a}^4\bigr>+90\,\bigl<{\bf a}^2\bigr>\,\bigl<{\bf a}^4\bigr>+144\,\bigl<{\bf a}\bigr>\,\bigl<{\bf a}^5\bigr>\nonumber\\
&&\qquad\left.-120\,\bigl<{\bf a}^6\bigr>\right]\,,
\label{16}
\end{eqnarray}

\noindent etc., where $\bigl<\cdot\bigr>={\bf tr}(\cdot)$. Discriminants of a square matrix are not so widely known objects. They appear in the literature under different names or indirectly in several contexts.

The discriminants (\ref{16}) satisfy the remarkable recurrence relation

\begin{equation}
{{\partial c_s({\bf a})}\over{\partial{\bf a}}}\,\cdot\,{\bf a}-c_s({\bf a})\,{\bf I}=-
{{\partial c_{s+1}({\bf a})}\over{\partial{\bf a}}}\,.
\label{17}
\end{equation}

\noindent where $(\partial c_s({\bf a})/\partial{\bf a})$ is the matrix with components

\begin{equation}
\left({{\partial c_s({\bf a})}\over{\partial{\bf a}}}\right)_{ij}
={{\partial c_s({\bf a})}\over{\partial a_{ij}}}
\label{18}
\end{equation}

We can state the Cayley--Hamilton theorem as:

\bigskip

\noindent {\bf Theorem.} (Cayley--Hamilton) {\it A $d$--dimensional matrix ${\bf a}$ satisfies}

\begin{equation}
{{\partial c_d({\bf a})}\over{\partial{\bf a}}}\,\cdot\,{\bf a}-c_d({\bf a})\,{\bf I}\equiv0\,.
\label{19}
\end{equation}

\bigskip

\noindent This result follows from (\ref{17}), for $s=d$, reminding that $c_{d+1}({\bf a})\equiv0$. For the first values of $d$ the explicit statement of the Cayley--Hamilton theorem is

\begin{eqnarray}
{\bf a}\,-c_1({\bf a})\,{\bf I}&\equiv&0\,,\nonumber\\
{\bf a}^2-c_1({\bf a})\,{\bf a}+c_2({\bf a})\,{\bf I}&\equiv&0\,,\nonumber\\
{\bf a}^3-c_1({\bf a})\,{\bf a}^2+c_2({\bf a})\,{\bf a}-c_3({\bf a})\,{\bf I}&\equiv&0\,,
\nonumber\\
{\bf a}^4-c_1({\bf a})\,{\bf a}^3+c_2({\bf a})\,{\bf a}^2-c_3({\bf a})\,{\bf a}
+c_4({\bf a})\,{\bf I}&\equiv&0\,,
\label{20}
\end{eqnarray}

\noindent etc.

According to the Cayley--Hamilton theorem, only the first $d$ powers of ${\bf a}$ are linearly independent. Therefore, only $d$ of the traces (\ref{09}) are algebraically independent. The ideal situation would be therefore to have a family of invariants such that only $d$ of them are non--trivial and algebraically independent. Discriminants have this property. For a $d$ dimensional matrix, only the first $d$ discriminants are non--trivial, while the discriminants of an order higher than $d$ are identically zero, $c_s({\bf a})\equiv0$, for $s>d$. This result is equivalent to the Cayley--Hamilton theorem. This reformulation of the Cayley--Hamilton theorem is possible because we can establish an equivalence between traces, power sums, elementary symmetric products and then $c_s\equiv0$. Therefore, we consider discriminants as our fundamental family of invariants.

If $c_d({\bf a})\not=0$, then from (\ref{19}) follows that there exist an inverse matrix ${\bf a}^{-1}$ which is given by

\begin{equation}
{\bf a}^{-1}={1\over{c_d({\bf a})}}\,{{\partial c_d({\bf a})}\over{\partial{\bf a}}}\,.
\label{21}
\end{equation}

\noindent In terms of components this inverse matrix is given by

\begin{equation}
({\bf a}^{-1})_{ij}={1\over{c_d({\bf a})}}\,{{\partial c_d({\bf a})}\over{\partial a_{ij}}}\,.
\label{22}
\end{equation}

We have two algorithms to compute the inverse of a matrix ${\bf a}$:

\begin{itemize}

\item[A1] The first one is based on (\ref{04}). In that case we have $n^2$ unknowns $(a^{-1})_{ij}$ and $n^2$ equations. The condition to have a solution is $\det({\bf a})\not=0$.

\item[A2] The second algorithm is based on the discriminant $c_d({\bf a})$. If $c_d({\bf a})\not=0$, then the inverse matrix is given by (\ref{21}).

\end{itemize}

\noindent If $\lambda_1,\cdots,\lambda_d$ are the eigenvalues of the matrix ${\bf a}$, then $c_d({\bf a})=\lambda_1\cdots\lambda_d$, which is the definition of the determinant of ${\bf a}$. Therefore, $c_d({\bf a})=\det({\bf a})$. This result explains why $c_d({\bf a})$ and $\det({\bf a})$, even when constructed with different algorithms lead to the same inverse matrix, namely (\ref{05}) and (\ref{21}). For simplicity we denote $a=\det({\bf a})=c_d({\bf a})$. Then, the equation (\ref{21}) is rewritten as

\begin{equation}
{\bf a}^{-1}={1\over a}\,{{\partial a}\over{\partial{\bf a}}}\,.
\label{23}
\end{equation}

\noindent For ordinary matrices, ${\bf a}$, the two previous algorithms are equivalent and yield the same inverse matrix ${\bf a}^{-1}$. However, for higher rank matrices only the second algorithm, based on (\ref{21}), admits a generalization.

\subsection{Matrices and tensors}

For the purpose of generalization to higher rank matrices it is convenient to represent matrices by means of tensors. We adopt the index notation of tensor calculus. In this case it is necessary to distinguish between covariant and contravariant indices. According to this scheme, a matrix ${\bf a}$ becomes represented by a second--rank covariant tensor ${\bf a}$ with components $a_{ij}$. Also we will need to consider contravariant tensors ${\bf g}$ with components $g^{ij}$. Furthermore, we adopt the summation convention according to which an index is summed up over its range if it appears once as a covariant index and once as a contravariant index; for example, $a_{ik}g^{kj}$ means $\sum_{k=1}^d a_{ik}g^{kj}$.

Let ${\bf a}$ be a covariant tensor with components $a_{ij}$. The unit element ${\bf e}$ for the tensor multiplication is a tensor with components ${e^i}_j$ given by

\begin{equation}
{e^i}_j=\delta^i_j=\cases{1\,,&for $i=j$;\cr0\,,&otherwise.\cr}
\label{24}
\end{equation}

\noindent The inverse ${\bf a}^{-1}$ is a tensor with components $(a^{-1})^{ij}$ such that

\begin{equation}
(a^{-1})^{ik}\,a_{jk}=\delta^i_j\,.
\label{25}
\end{equation}

\noindent Since the inverse tensor is a contravariant tensor the notation $(a^{-1})^{ij}$ becomes redundant; therefore, we simply write $a^{ij}$ for the components of the inverse tensor. For example, in term of components, equation (\ref{23}) becomes 

\begin{equation}
a^{ij}={1\over a}\,{{\partial a}\over{\partial a_{ij}}}\,,
\label{26}
\end{equation}

\noindent while equation (\ref{25}) becomes

\begin{equation}
a^{ik}\,a_{jk}=\delta^i_j\,.
\label{27}
\end{equation}

In order to define a multiplication operation in a tensor language it is necessary to consider a metric tensor ${\bf g}$ with components $g_{ij}$. If $g=\det({\bf g})\not=0$, then there exist an inverse ${\bf g}^{-1}$ with components $g^{ij}$. Then, if ${\bf a}$ and ${\bf b}$ are covariant tensors with components $a_{ij}$ and $b_{ij}$, the multiplication operation is defined by the product tensor ${\bf c}$ with components

\begin{equation}
c_{ij}=a_{ik}\,g^{lk}\,b_{lj}\,.
\label{28}
\end{equation}

\noindent The trace is defined as

\begin{equation}
{\bf tr}_{{\bf g}}({\bf a})=g^{ij}\,a_{ij}\,.
\label{29}
\end{equation}

\noindent The definition of the trace in terms of the same metric tensor ${\bf g}$ used for the multiplication is necessary in order to preserve the property of ciclicity of the trace.

In order to agree with (\ref{01}), it would be necessary to choose

\begin{equation}
g^{ij}=\cases{1\,,&if $i=j$;\cr0\,,&otherwise.\cr}
\label{30}
\end{equation}

\noindent However, this definition has not an invariant tensorial meaning and therefore it is better to keep working with a generic metric tensor ${\bf g}^{-1}$ with components $g^{ij}$ unrelated to (\ref{30}).

\subsection{Permutation products and discriminants}

For higher rank matrices neither a natural multiplication operation nor a natural extension of the concept of a trace do exist. Therefore, we need to elaborate a definition of discriminants without making reference to multiplications or traces, and, in such a way, that the resulting definition reduces to the usual definition for matrices of second rank and is easily generalizable to matrices of higher rank. To this purpose, we define permutation products in terms of ${\bf g}$. As a motivation and justification to do so we start by considering the definition in tensor calculus of the determinant of a tensor ${\bf a}$.

The Levi--Civita symbol is defined by

\begin{equation}
\epsilon^{i_1\cdots i_d}=\cases{\quad1& if $i_1\cdots i_d$ is an even permutation of $1\cdots d$;\cr-1& if $i_1\cdots i_d$ is an odd permutation of $1\cdots d$;\cr\quad0&if $i_1\cdots i_d$ is not a permutation of $1\cdots d$.}
\label{31}
\end{equation}

\noindent Then, the determinant of a tensor ${\bf a}$ with components $a_{ij}$ is given by

\begin{equation}
\det({\bf a})={1\over{d!}}\,\epsilon^{i_1\cdots i_d}\,\epsilon^{j_1\cdots j_d}\,a_{i_1j_1}\,
\cdots\,a_{i_d j_d}\,.
\label{32}
\end{equation}

\noindent Substituting (\ref{32}) in (\ref{26}) we obtain the explicit expression

\begin{equation}
a^{ij}={1\over{(d-1)!}}\,{1\over a}\,\epsilon^{i i_1\cdots i_{(d-1)}}\,
\epsilon^{j j_1\cdots j_{(d-1)}}\,a_{i_1j_1}\,\cdots\,a_{i_{(d-1)}j_{(d-1)}}\,.
\label{33}
\end{equation}

\noindent We can verify that $a^{ij}$ so defined satisfies (\ref{27}). It can also be verified that equation (\ref{33}) is equation (\ref{05}) written in terms of components.

Therefore, for second rank covariant tensors the discriminants can be constructed with the use of ${\bf g}$ for products and traces, as in (\ref{28}) and (\ref{29}), and the definitions for discriminants in (\ref{16}). However, for the determinant, which corresponds to the discriminant $c_d({\bf a})$, we have (\ref{32}) which involves ${\bf a}$ alone and that makes no reference to ${\bf g}$. In order to avoid, and explain, this duplicity we develope a unified scheme to construct discriminants and which is well suited for the generalization to higher rank matrices. 

Let us consider a metric tensor ${\bf g}$ with componentes $g_{ij}$ and define the permutation tensors ${\bf q}$ by

\begin{equation}
q_s^{i_1j_1\cdots i_s j_s}({\bf g})={1\over{s!(d-s)!}}\,{1\over g}\,\epsilon^{i_1\cdots i_s i_{s+1}\cdots i_d}\,\epsilon^{j_1\cdots j_s j_{s+1}\cdots j_d}\,g_{i_{s+1}j_{s+1}}\,\cdots\,
g_{i_d j_d}\,.
\label{34}
\end{equation}

\noindent The tensors ${\bf q}$ are non trivial only for $s\leq d$, and ${\bf q}_s\equiv0$ for $s>d$. We define the discriminants $c_s^{{\bf g}}({\bf a})$ for a tensor ${\bf a}$ by

\begin{equation}
c_s^{{\bf g}}({\bf a})=q_s^{i_1j_1\cdots i_s j_s}({\bf g})\,a_{i_1j_1}\,\cdots\,a_{i_s j_s}\,.
\label{35}
\end{equation}

\noindent Then, the discriminants $c_s^{{\bf g}}({\bf a})$ are given by

\begin{equation}
c_s^{{\bf g}}({\bf a})={1\over{s!(d-s)!}}\,{1\over g}\,
\epsilon^{i_1\cdots i_s i_{s+1}\cdots i_d}\,\epsilon^{j_1\cdots j_s j_{s+1}\cdots j_d}\,
a_{i_1j_1}\,\cdots\,a_{i_s j_s}\,g_{i_{s+1}j_{s+1}}\,\cdots\,g_{i_d j_d}\,.
\label{36}
\end{equation}

For second--rank tensors, the case under consideration, the permutation tensors (\ref{34}) can be rewritten as

\begin{equation}
q_s^{i_1j_1\cdots i_s j_s}({\bf g})={1\over{s!}}\,g^{|[i_1j_1}\,\cdots\,g^{i_s j_s]|}\,,
\label{37}
\end{equation}

\noindent where $|[\cdots]|$ denotes complete antisymmetry with respect to the indices $j$'s or, equivalently, with respect to the indices $i$'s. For the first values of $s$ the tensors ${\bf q}$ are given by

\begin{eqnarray}
q_1^{ij}({\bf g})&=&g^{ij}\,,\nonumber\\
q_2^{i_1j_1i_2j_2}({\bf g})&=&{1\over2}\,(g^{i_1j_1}\,g^{i_2j_2}-g^{i_1j_2}\,g^{i_2j_1})\,,
\nonumber\\
q_3^{i_1j_1i_2j_2i_3j_3}({\bf g})&=&{1\over{3!}}\,[g^{i_1j_1}\,g^{i_2j_2}\,g^{i_3j_3}
-(g^{i_1j_1}\,g^{i_2j_3}\,g^{i_3j_2}\nonumber\\
&&\qquad+g^{i_1j_3}\,g^{i_2j_2}\,g^{i_3j_1}+g^{i_1j_2}\,g^{i_2j_1}\,g^{i_3j_3})\nonumber\\
&&\qquad+(g^{i_1j_2}\,g^{i_2j_3}\,g^{i_3j_1}+g^{i_1j_3}\,g^{i_2j_1}\,g^{i_3j_2})]\,,
\label{38}
\end{eqnarray}

\noindent etc. When we restrict ${\bf g}$ to (\ref{30}), (\ref{36}) reduces to the relations (\ref{16}) with $\bigl<\cdot\bigr>={\bf tr}_{{\bf g}}(\cdot)$; therefore this is a valid generalization of the discriminants.

We can verify that

\begin{equation}
{{\partial(g\,c^{{\bf g}}_s({\bf a}))}\over{\partial{\bf g}}}=
{{\partial(g\,c^{{\bf g}}_{s+1}({\bf a}))}\over{\partial{\bf a}}}\,,
\label{39}
\end{equation}

\noindent and from here it follows that

\begin{equation}
{{\partial c^{{\bf g}}_s({\bf a})}\over{\partial{\bf g}}}+c^{{\bf g}}_s({\bf a})\,{\bf g}^{-1}=
{{\partial c^{{\bf g}}_{s+1}({\bf a})}\over{\partial{\bf a}}}\,.
\label{40}
\end{equation}

\noindent This is a recurrence relation analogous to (\ref{17}).

We can now reformulate the Cayley--Hamilton theorem as:

\bigskip

\noindent {\bf Theorem.} (Cayley--Hamilton) {\it A $d$--dimensional tensor ${\bf a}$ satisfies}

\begin{equation}
{{\partial c^{{\bf g}}_d({\bf a})}\over{\partial{\bf g}}}
+c^{{\bf g}}_d({\bf a})\,{\bf g}^{-1}\equiv0\,.
\label{41}
\end{equation}

\bigskip

\noindent This results follows from (\ref{40}), for $s=d$, reminding that $c^{{\bf g}}_{d+1}({\bf a})\equiv0$.

There are two particularly interesting instances of relation (\ref{41}):

\begin{itemize}

\item[B1] The first case appears if we insist on ${\bf g}^{-1}$ as given in (\ref{30}) in an attempt to reproduce the standard results. Then, the relations in (\ref{41}) reduce to the relations of matrix calculus given by (\ref{20}).

\item[B2] The second case appears if we want to have expressions concomitant of ${\bf a}$ alone. In this case the only possible choice is ${\bf g}={\bf a}$. However, all relations in (\ref{41}) collapse to useless identities.

\end{itemize}

\noindent The generalization of the first case to higher rank matrices is excluded since it is not possible to define a metric tensor ${\bf g}$ similar to an identity or unit matrix in an invariant way. On the other hand, the generalization of the second case is possible and is the one which gives rise to discriminants and the Cayley--Hamilton theorem for higher--rank matrices.

\bigskip

The componentes of ${\bf q}_d$ are completely antisymmetric in the $d$ indices $i$ and the $d$ indices $j$. Therefore, they must be the product of Levi--Civita symbols in those indices

\begin{equation}
q_d^{i_1\cdots i_d j_1\cdots j_d}({\bf g})={1\over{d!}}\,g^{|[i_1j_1}\,\cdots\,g^{i_d j_d]|}
={1\over g}\,{1\over{d!}}\,\epsilon^{i_1\cdots i_d}\,\epsilon^{j_1\cdots j_d}\,.
\label{42}
\end{equation}

\noindent  Then, we obtain

\begin{equation}
c_d^{{\bf g}}({\bf a})={1\over g}\,{1\over{d!}}\,\epsilon^{i_1\cdots i_d}\,
\epsilon^{j_1\cdots j_d}\,a_{i_1j_1}\,\cdots\,a_{i_d j_d}={a\over g}\,.
\label{43}
\end{equation}

\noindent Let us denote $a_g=c_d^{{\bf g}}({\bf a})=a/g$. If $a\not=0$ and $g\not=0$, we obtain

\begin{equation}
{\bf a}^{-1}={1\over a}\,{{\partial a}\over{\partial{\bf a}}}={1\over{a_g}}\,{{\partial a_g}
\over{\partial{\bf a}}}\,.
\label{44}
\end{equation}

\noindent This relation explains why the determinant, as defined in (\ref{32}) (which does not involves ${\bf g}$), gives the same inverse tensor as $c_d^{{\bf g}}({\bf a})$.

\section{Higher rank matrices}

The concept of a higher rank matrix, and the corresponding determinant, was introduced by Cayley in 1845 \cite{Ca} and it was later developed by Schl\"afli \cite{Sc} and by Pascal \cite{Pa}. More recently, matrices of higher rank have been studied in \cite{GKZ1,GKZ2,GKZ3,WE,WZ1,WZ2}. Particularly interesting are \cite{GKZ2} and \cite{GKZ3} where a general account on the subject, with many generalizations and applications, can be found. The interested reader may consult these references for further detail.

For higher--rank matrices there is not a natural multiplication operation in the sense that the product of two higher rank matrices be a matrix of the same rank and covariance. Therefore, the construction of discriminants must be done by generalizing a definition of discriminants which do not involve any multiplication operation. The definition of discriminants in terms of permutation products satisfies this requirement, but even rank and odd rank cases must be considered separately. For the even rank case we take the fourth--rank case as a prototype. Then, we construct discriminants, the determinant, and show that a polynomial relation among discriminants and products of the higher rank matrix vanishes identically, which is a statement similar to the Cayley--Hamilton theorem. This algorithm can be easily extended to matrices of any arbitrary even rank. For the odd rank case we consider third--rank matrices as a prototype and show that in this case it is necessary to introduce an even rank matrix constructed as the direct product of the odd rank matrix. Then, discriminants, the determinant and the Cayley--Hamilton theorem follow as for the even rank case.

Let us remind once again that higher--rank matrices will be represented by higher--rank tensors. Therefore, we will talk, mostly, of tensors rather than matrices.

\subsection{The even rank case}

We consider fourth--rank matrices as a representative of even--rank matrices; all other even--rank cases can be dealt with in a similar way. 

The simplest extensions of the concept of determinant to matrices of higher rank is the one due to Cayley, which is a direct generalisation of the Leibniz formula; namely

\begin{equation}
{\det}_C(A)=\sum_{\pi_2\cdots\pi_r\in S_d}\,{\rm sign}(\pi_2\,\cdots\,\pi_r)\,\left(\prod_{i=1}
^d\,A_{i\pi_2(i)\cdots\pi_r(i)}\right)\,.
\label{45}
\end{equation}

\noindent Then, the determinant of a fourth--rank tensor ${\bf A}$ with components $A_{ijkl}$ is defined as a direct generalization of (\ref{32}), namely,

\begin{equation}
\det({\bf A})={1\over{d!}}\,\epsilon^{i_1\cdots i_d}\,\cdots\,\epsilon^{l_1\cdots l_d}\,
A_{i_1j_1k_1l_1}\,\cdots\,A_{i_d j_d k_d l_d}\,.
\label{46}
\end{equation}

\noindent As before we simplify the notation by writing $A=\det({\bf A})$. In analogy with (\ref{23}) we define

\begin{equation}
{\bf A}^{-1}={1\over A}\,{{\partial A}\over{\partial{\bf A}}}\,.
\label{47}
\end{equation}

\noindent In term of components

\begin{equation}
A^{ijkl}={1\over A}\,{{\partial A}\over{\partial A_{ijkl}}}\,.
\label{48}
\end{equation}

\noindent Then

\begin{equation}
A^{ijkl}={1\over{(d-1)!}}\,{1\over A}\,\epsilon^{i i_1\cdots i_{(d-1)}}\,\cdots\,
\epsilon^{l l_1\cdots l_{(d-1)}}\,A_{i_1j_1k_1l_1}\,\cdots\,A_{i_{(d-1)}j_{(d-1)}k_{(d-1)}l_{(d-1)}}\,.
\label{49}
\end{equation}

\noindent This higher rank tensor satisfies,

\begin{equation}
A^{i k_1k_2k_3}\,A_{j k_1k_2k_3}=\delta^i_j\,,
\label{50}
\end{equation}

\noindent which is a relation similar to (\ref{27}). The definitions in (\ref{46}) and (\ref{49}) were used in previous works \cite{Ta93,TRMC,TR} concerning the application of fourth--rank geometry in the formulation of an alternative theory for the gravitational field.

\bigskip

As an example let us consider the simple case $d=2$. The determinant (\ref{46}) is given by

\begin{equation}
A=A_{0000}\,A_{1111}-4\,A_{0001}\,A_{0111}+3\,A_{0011}^2\,.
\label{51}
\end{equation}

\noindent The components of the inverse matrix ${\bf A}^{-1}$ are given by

\begin{eqnarray}
A^{0000}&=&\quad{1\over A}\,A_{1111}\,,\nonumber\\
A^{0001}&=&-{1\over A}\,A_{0111}\,,\nonumber\\
A^{0011}&=&\quad{1\over A}\,A_{0011}\,.
\label{52}
\end{eqnarray}

\noindent and similar expressions for the other components. In order to verify the validity of equation (\ref{50}) let us consider the cases $(00)$ and $(01)$. We obtain

\begin{eqnarray}
A^{0ijk}\,A_{0ijk}&=&1\,,\nonumber\\
A^{0ijk}\,A_{1ijk}&=&0\,,
\label{53}
\end{eqnarray}

\noindent and similar relations for the other indices.

\bigskip

In a way similar to (\ref{34}) we define the permutation tensors ${\bf Q}$ by

\begin{eqnarray}
Q_s^{i_1j_1k_1l_1\cdots i_s j_s k_s l_s}({\bf G})&=&{1\over{s!(d-s)!}}\,{1\over G}\,
\epsilon^{i_1\cdots i_s i_{s+1}\cdots i_d}\,\epsilon^{j_1\cdots j_s j_{s+1}\cdots j_d}
\nonumber\\
&&\qquad\epsilon^{k_1\cdots k_s k_{s+1}\cdots k_d}\,\epsilon^{l_1\cdots l_s l_{s+1}\cdots l_d}
\nonumber\\
&&\qquad G_{i_{s+1}j_{s+1}k_{s+1}l_{s+1}}\,\cdots\,G_{i_d j_d k_d l_d}\,,
\label{54}
\end{eqnarray}

\noindent where $G=\det({\bf G})$. The tensors ${\bf Q}$ are non trivial only for $s\leq d$, and ${\bf Q}_s\equiv0$ for $s>d$. Then, we define the discriminants $C_s^{{\bf G}}({\bf A})$ by

\begin{equation}
C_s^{{\bf G}}({\bf A})=Q_s^{i_1j_1k_1l_1\cdots i_s j_s k_s l_s}({\bf G})\,A_{i_1j_1k_1l_1}\,
\cdots\,A_{i_s j_s k_s l_s}\,.
\label{55}
\end{equation}

\noindent Explicitly

\begin{eqnarray}
C_s^{{\bf G}}({\bf A})={1\over{s!(d-s)!}}\,{1\over G}\,
\epsilon^{i_1\cdots i_s i_{s+1}\cdots i_d}\,\epsilon^{j_1\cdots j_s j_{s+1}\cdots j_d}\,
\epsilon^{k_1\cdots k_s k_{s+1}\cdots k_d}\,\epsilon^{l_1\cdots l_s l_{s+1}\cdots l_d}
\nonumber\\
\times\,A_{i_1j_1k_1l_1}\,\cdots\,A_{i_s j_s k_s l_s}\,G_{i_{s+1}j_{s+1}k_{s+1}l_{s+1}}\,\cdots
\,G_{i_d j_d k_d l_d}\,.
\label{56}
\end{eqnarray}

\noindent Then, in a way similar to (\ref{40}), it is possible to verify that the discriminants $C_s^{{\bf G}}({\bf A})$ satisfy the recurrence relation

\begin{equation}
{{\partial C^{{\bf G}}_s({\bf A})}\over{\partial{\bf G}}}+C^{{\bf G}}_s({\bf A})\,{\bf G}^{-1}
={{\partial C^{{\bf G}}_{s+1}({\bf A})}\over{\partial{\bf A}}}\,.
\label{57}
\end{equation}

\noindent Then, we have

\bigskip

\noindent {\bf Theorem.} (Cayley--Hamilton) {\it A $d$--dimensional tensor ${\bf A}$ of fourth--rank satisfies}

\begin{equation}
{{\partial C^{{\bf G}}_d({\bf A})}\over{\partial{\bf G}}}
+C^{{\bf G}}_d({\bf A})\,{\bf G}^{-1}\equiv0\,.
\label{58}
\end{equation}

\bigskip

The components of ${\bf Q}_d$ are completely symmetric in indices $i$, $j$, $k$, and $l$; therefore, they must be the product of Levi--Civita symbols. We then have

\begin{equation}
Q_d^{i_1j_1k_1l_1\cdots i_d j_d k_d l_d}({\bf G})={1\over G}\,{1\over{d!}}\,\epsilon^{i_1\cdots i_d}\,\cdots\,\epsilon^{l_1\cdots l_d}\,,
\label{59}
\end{equation}

\noindent which is the result similar to (\ref{42}). Therefore

\begin{equation}
C_d^{{\bf G}}({\bf A})={A\over G}\,.
\label{60}
\end{equation}

\noindent Let us denote $A_G=C_d^{{\bf G}}({\bf A})=A/G$. If $A\not=0$ and $G\not=0$, we obtain

\begin{equation}
{\bf A}^{-1}={1\over A}\,{{\partial A}\over{\partial{\bf A}}}={1\over{A_G}}\,{{\partial A_G}
\over{\partial{\bf A}}}\,,
\label{61}
\end{equation}

\noindent which is the relation similar to (\ref{44}).

There is one particularly interesting instance of the relations (\ref{58}). If we want to have expressions concomitant of ${\bf A}$ alone, then the only possible choice is ${\bf G}={\bf A}$. To this purpose it is useful to observe that if $s=d$, and only in this case, the permutation tensors (\ref{54}) can be written in a way similar to (\ref{37}), namely

\begin{equation}
Q_d^{i_1j_1k_1l_1\cdots i_d j_d k_d l_d}({\bf G})={1\over{d!}}\,G^{|[i_1j_1k_1l_1}\,\cdots\,
G^{i_d j_d k_d l_d]|}\,.
\label{62}
\end{equation}

\bigskip

As an example of the results above let us consider the case $d=2$. For the tensor ${\bf Q}_2$, we obtain

\begin{eqnarray}
Q_2^{i_1j_1k_1l_1i_2j_2k_2l_2}({\bf G})&=&{1\over2}\,[G^{i_1j_1k_1l_1}\,G^{i_2j_2k_2l_2}
-(G^{i_1j_1k_1l_2}\,G^{i_2j_2k_2l_1}\nonumber\\
&&\qquad+G^{i_1j_1k_2l_1}\,G^{i_2j_2k_1l_2}+G^{i_1j_2k_1l_1}\,G^{i_2j_1k_2l_2}\nonumber\\
&&\qquad+G^{i_2j_1k_1l_1}\,G^{i_1j_2k_2l_2})+(G^{i_1j_1k_2l_2}\,G^{i_2j_2k_1l_1}\nonumber\\
&&\qquad+G^{i_1j_2k_1l_2}\,G^{i_2j_1k_2l_1}+G^{i_2j_1k_1l_2}\,G^{i_1j_2k_2l_1})]\,
\label{63}
\end{eqnarray}

\noindent The determinant is given by

\begin{eqnarray}
C^{{\bf G}}_2({\bf A})&=&{1\over2}\,[(G^{ijkl}\,A_{ijkl})^2-4\,G^{ijkl}\,A_{jklm}\,
G^{mnpq}\,A_{npqi}\nonumber\\
&&\qquad+3\,G^{ijkl}\,A_{klmn}\,G^{mnpq}\,A_{pqij}]\,,
\label{64}
\end{eqnarray}

\noindent Finally, the corresponding polynomial identity is given by

\begin{eqnarray}
(G^{mnpq}\,A_{mnpq})\,A_{ijkl}-4\,A_{(i|mnp}\,G^{mnpq}\,A_{q|jkl)}&&\nonumber\\
+3\,A_{(ij|mn}\,G^{mnpq}\,A_{pq|kl)}-C_2^{{\bf G}}({\bf A})\,G_{ijkl}&=&0\,.
\label{65}
\end{eqnarray}

\noindent If we choose ${\bf G}={\bf A}$ the relation (\ref{65}) reduces to

\begin{equation}
A_{(ij|mn}\,A^{mnpq}\,A_{pq|kl)}-{1\over2}\,(A^{mnpq}\,A_{pqrs}\,A^{rstu}\,A_{tumn})\,
A_{ijkl}\equiv0\,,
\label{66}
\end{equation}

\noindent where $(\cdot)$ means that the enclosed indices are symmetrised and $|\cdot|$ means that the enclosed indices are excluded from the symmetrisation.

\subsection{The odd rank case}

We consider third--rank matrices as a representative of odd--rank matrices; all other odd--rank cases can be dealt with in a similar way. 

Let ${\bf s}$ be a third--rank tensor with components $s_{ijk}$. A naive definition of the determinant would be the natural generalization of (\ref{32}) and (\ref{46}), namely

\begin{equation}
\det({\bf s})={1\over{d!}}\,\epsilon^{i_1\cdots i_d}\,\cdots\,\epsilon^{k_1\cdots k_d}
\,s_{i_1j_1k_1}\,\cdots\,s_{i_d j_d k_d}\,.
\label{67}
\end{equation}

\noindent However, it is possible to verify that

\begin{equation}
\epsilon^{i_1\cdots i_d}\,\cdots\,\epsilon^{k_1\cdots k_d}\,s_{i_1j_1k_1}\,\cdots
\,s_{i_d j_d k_d}\equiv0\,.
\label{68}
\end{equation}

\noindent This result is due to the odd number of $\epsilon$'s in (\ref{67}) which add all contributions to zero. 

In order to obtain some indication as to the correct way to define a determinant for odd--rank tensors, let us consider the simple case of a completely symmetric third--rank tensor ${\bf s}$ with components $s_{ijk}$ in dimension $d$. Then, let us look for an inverse tensor ${\bf s}^{-1}$ with components $s^{ijk}$ such that a relation similar to (\ref{27}) and (\ref{50}) holds, namely

\begin{equation}
s^{ikl}\,s_{jkl}=\delta^i_j\,.
\label{69}
\end{equation}

\noindent The number of unknowns in (\ref{69}) is $d(d+1)(d+2)/6$ while the number of equations is $d^2$. This algebraic system of equations is underdetermined, except for $d=2$. In this last case the solution is given by

\begin{eqnarray}
s^{000}&=&{1\over{s^2}}\,\left(s_{000}\,s_{111}^2+2\,s_{011}^3-3\,s_{001}\,s_{011}\,s_{111}
\right)\,,\nonumber\\
s^{001}&=&{1\over{s^2}}\,\left(2\,s_{111}\,s_{001}^2-s_{000}\,s_{011}\,s_{111}-s_{001}\,
s_{011}^2\right)\,,
\label{70}
\end{eqnarray}

\noindent and similar relations for the other components, where

\begin{equation}
s^2=s_{000}^2\,s_{111}^2-6\,s_{000}\,s_{001}\,s_{011}\,s_{111}+4\,s_{000}\,s_{011}^3
+4\,s_{111}\,s_{001}^3-3\,s_{001}^2\,s_{011}^2\,.
\label{71}
\end{equation}

\noindent Let us observe that

\begin{eqnarray}
s^{000}&=&{1\over{2s^2}}\,{{\partial s^2}\over{\partial s_{000}}}={1\over s}\,{{\partial s}\over{\partial s_{000}}}\,,\nonumber\\
s^{001}&=&{1\over3}\,{1\over{2s^2}}\,{{\partial s^2}\over{\partial s_{001}}}={1\over3}\,{1\over s}\,{{\partial s}\over{\partial s_{001}}}\,,
\label{72}
\end{eqnarray}

\noindent leading to

\begin{equation}
s^{ijk}={1\over{2s^2}}\,{{\partial s^2}\over{\partial s_{ijk}}}={1\over s}\,{{\partial s}\over{\partial s_{ijk}}}\,.
\label{73}
\end{equation}

\noindent Therefore, the role of the determinant, in a way similar as appears in (\ref{26}) and (\ref{48}), is played by $s$. 

For any matrix ${\bf S}$ the product of $\epsilon$'s and ${\bf S}$'s is a meaningful quantity only for an expression in which ${\bf S}$ is an even rank tensor. Then $s^2$ must be related to the determinant of some higher even rank tensor ${\bf S}$, and this determinant must be a quadratic expression (since $d=2$) in this higher rank tensor ${\bf S}$. Since $s^2$ is a quartic expression in ${\bf s}$ we have that ${\bf S}$ must be quadratic in ${\bf s}$. A solution satisfying the requirements above is given by

\begin{equation}
S_{i_1j_1k_1i_2j_2k_2}=s_{(i_1j_1k_1}\,s_{i_2j_2k_2)}\,.
\label{74}
\end{equation}

\noindent For $d=2$, we obtain

\begin{eqnarray}
S_{000000}&=&s_{000}^2\,,\nonumber\\
S_{000001}&=&s_{000}\,s_{001}\,,\nonumber\\
S_{000011}&=&{1\over5}\,(2\,s_{000}\,s_{011}+3\,s_{001}^2)\,,\nonumber\\
S_{000111}&=&{1\over{19}}\,(s_{000}\,s_{111}+9\,s_{001}\,s_{011})\,,
\label{75}
\end{eqnarray}

\noindent and similar relations for the other components. 

The determinant of a sixth--rank tensor ${\bf S}$ with components $S_{i_1i_2i_3i_4i_5i_6}$ is defined through a direct extension of the definition in (\ref{46}). We obtain

\begin{equation}
S={1\over{d!}}\,\epsilon^{i_1\cdots i_d}\,\cdots\,\epsilon^{n_1\cdots n_d}\,
S_{i_1\cdots n_1}\,\cdots\,S_{i_d\cdots n_d}\,.
\label{76}
\end{equation}

\noindent In the the two--dimensional case this determinant is given by

\begin{equation}
S=S_{000000}\,S_{111111}-6\,S_{000001}\,S_{111110}+15\,S_{000011}\,S_{111100}-10\,S_{000111}^2\,.
\label{77}
\end{equation}

\noindent If we replace (\ref{75}) in (\ref{77}) we obtain, up to an irrelevant multiplicative constant, the expression (\ref{71}).

Therefore, for odd rank matrices the recipe is to construct an even rank matrix as the direct product of the original matrix. Then, the construction of invariants proceeds as for the even rank case.

\section{Concluding remarks}

We have developed an algorithm to construct invariants for higher rank matrices. We constructed determinants and exhibit an extension of the Cayley--Hamilton theorem to higher rank matrices.

Among the possible applications of the present work we have the quantum mechanics of entangled states. Indeed, in order to obtain a measurement of entanglement one needs to construct invariants associated to higher rank matrices; see \cite{ALP,BLT,BLTV,CKW,GRB,Le,LP,LT03,LT06,Mi} for several interesting results.

Higher rank tensors, which look similar to higher rank matrices, appear in several contexts such as in Finsler geometry \cite{As,Ru} and in fourth--rank gravity \cite{Ta93,TRMC,TR}. The results presented here are a first step for the construction of differential invariants for higher--rank tensors. 

\section*{Acnowledgements}

This work was partially done at the {\bf Abdus Salam} International Centre for Theoretical Physics, Trieste.


\end{document}